\documentclass[12pt,preprint]{aastex}

\shorttitle{PWN of PSR J2021$+$3651}
\shortauthors{}

\begin{document}

\title{Rings and Jets around PSR J2021+3651: the `Dragonfly Nebula'}

\author{Adam Van Etten\altaffilmark{1}, Roger W. Romani\altaffilmark{1} \& C.-Y. Ng\altaffilmark{2}}
\altaffiltext{1}{Department of Physics, Stanford University, Stanford, CA 94305}
\altaffiltext{2}{School of Physics, University of Sydney, NSW 2006, Australia}
\email{ave@stanford.edu, rwr@astro.stanford.edu, ncy@physics.usyd.edu.au}

\begin{abstract}
We describe recent Chandra ACIS observations of the Vela-like pulsar
PSR J2021+3651 and its pulsar wind nebula (PWN).  This `Dragonfly Nebula'
displays an axisymmetric morphology, with bright inner jets, a double-ridged
inner nebula, and a $\sim30^{\prime\prime}$ polar jet. The PWN is
embedded in faint diffuse emission: a bow shock-like structure with
standoff $\sim1^{\prime}$ brackets the pulsar to the east and emission trails off
westward for 3-4\arcmin.
Thermal (k$T_{\infty}=0.16\pm0.02$ keV) and power law emission are detected
from the pulsar. The nebular X-rays show spectral steepening from
$\Gamma=1.5$ in the equatorial torus to $\Gamma=1.9$ in the outer nebula, suggesting
synchrotron burn-off. A fit to the `Dragonfly' structure suggests a
large (86$\pm1^{\circ}$)
inclination with a double equatorial torus. Vela is currently the only other
PWN showing such double structure. The $>$12\,kpc distance implied by the
pulsar dispersion measure is not supported by the X-ray data; spectral, scale and
efficiency arguments suggest a more modest 3-4\,kpc.
\end{abstract}


\section{Introduction}

        Pulsars inject a relativistic particle wind into their surroundings,
tapping their rotational kinetic energy.  This energetic wind is concentrated
toward the spin equator and, shocking against the surrounding medium, causes
particle acceleration and pitch angle scattering that results in a synchrotron
nebula \citep{rg74,kc84}.  Approximately 50 pulsars have an associated PWN,
and the polar jet/equatorial torus morphology is now seen to be quite common \citep{kp08}.
Many of the mechanisms governing the energetics and evolution of pulsar/PWN systems
are still unknown, such as how rotational kinetic energy is converted to particle
outflow, the manner in which polar jets are confined, and the details of
charge acceleration. 
The structure of PWNe offer insights into these puzzles, and also probe pulsar geometry,
relativistic shocks, and the properties of the ambient medium.  \object{PSR J2021+3651}
(hereafter J2021) is interesting for its highly structured PWN, and is also a
likely counterpart of the EGRET gamma ray source GEV 2020+3658.  Pulsars detected
in GeV gamma rays are currently of great interest with the successful launch of
AGILE and the imminent launch of GLAST.

Timing observations of this 103.7 ms pulsar show that J2021 is young
and energetic [characteristic age $\tau_c = {P/(2{\dot P})} = 1.7 \times 10^4$\,yr
and spindown luminosity ${\dot E} = 4 {\pi}^2 I{\dot P}{P^{-3}} = 3.4
\times 10^{36}$\,ergs\,s$^{-1}$ ], despite being rather radio
faint with a 1.4\,GHz flux density S$_{1400}$ $\approx 0.1$\,mJy \citep{ret02,het04}.
The dispersion measure of 369 ${\rm cm^{-3}}$ pc places it $\sim$12 kpc away on the far
edge of the outer spiral arm, according to the NE2001 electron model of
the galaxy \citep{cl02}.  Although its spin characteristics are similar
to those of other $\gamma$-ray pulsars [\object{Vela}: $d\approx300$ pc
\citep{det03}, $P$ = 89.3 ms, ${\dot E}$ = 6.9$\times10^{36}$ ergs\,s$^{-1}$;
\object{B1706--44}: $d\sim$3 kpc \citep{ret05}, $P$ = 103 ms,
${\dot E}$ = 3.4$\times10^{36}$ ergs\,s$^{-1}$], the
12 kpc distance assumed from the dispersion measure would make it
extraordinarily $\gamma$-ray efficient, as discussed by \citet{ret02}.

\citet{rrk01} noted the possible connection between GEV J2020+3658 and ASCA source
AX J2021.1+3651.  The ASCA data revealed spatially unresolved emission extending $\sim$8'
east to west and $\sim$4' north to south, with a hydrogen column density of
$N_{\rm H}$ = 5.0$\pm2.5\times 10^{21} {\rm cm^{-2}}$, and a power-law spectrum with
$\Gamma$=1.73$_{-0.28}^{+0.26}$.  The following year a targeted radio search by
\citet{ret02} discovered J2021; these authors reanalyzed the ASCA data and concluded that
a thermal component (k$T\sim$0.1 keV) in addition to the absorbed power law improved the
fit to the data, giving an $N_{\rm H}$ = 7.6$_{-3.5}^{+4.7} \times 10^{21} {\rm cm^{-2}}$.
\citet{het04} were able to resolve the inner equatorial PWN and jets surrounding J2021 with a
19 ks \emph{Chandra} ACIS-S observation, noting a toroidal morphology and possible double
structure; this observation had a sub-framed field of view and could not probe
the faint larger-scale structure. These authors report an 
$N_{\rm H}$ of 7.8$_{-1.4}^{+1.7} \times 10^{21} {\rm cm^{-2}}$,
a power law index of $\Gamma$=1.73$_{-0.2}^{+0.3}$ in the extended emission. 
Thermal emission from the pulsar at k$T_{\infty}$=0.15$\pm0.02$ keV was noted, along with
a possible pulse fraction of 65$\%$ from the pulsar, presumably mostly due to thermal emission.
We describe here a deeper \emph{Chandra}
exposure, which better resolves the innermost structure and maps the outer portions of the
nebula. The improved constraints on spectral and morphological parameters of the
PWN afforded by this deeper observation can help address the question of the
pulsar's distance and will be helpful in interpreting future high energy observations.

\section{Observations and Data Analysis}

We observed J2021 with \emph{CXO} in the ACIS-S configuration (chips 2-3 and 5-8),
with the pulsar positioned near the standard aim-point on the S3 chip during two epochs:
December 25-26 2006 with a total live time of 33.8 ks (obsID
\dataset[ADS/Sa.CXO#obs/8502]{8502} = `obs1')
and December 29-30 2006 for 59.4 ks (obsID \dataset[ADS/Sa.CXO#obs/7603]{7603} = `obs2').
All figures shown below are merged images of the two observations.  The CCDs
were operated in VFAINT mode to improve rejection of particle backgrounds.
Neither observation exhibited significant flaring, so we include all the data
in the analysis, for a total live time of 93.2 ks. These data are also
compared with an archival 19.0 ks observation on chips 6-8 taken
on February 12 2003 (obsID \dataset[ADS/Sa.CXO#obs/3901]{3901} = `obs0').
For that exposure, the chips were
windowed to 1/4 frame, covering only the bright central region of the PWN,
as described in \citet{het04}.

The data were analyzed using CIAO 3.4.0 and CALDB 3.4.0. 
We started with level=1 event files, removed pixel randomization,
applied time-dependent gain and CTI corrections, and implemented a background
cleaning algorithm for VFAINT mode which uses the outer 16 pixels
of the 5$\times$5 event island to improve discrimination between good events
and likely cosmic rays.  HEASARC's WebPIMMS
tool estimates a small pileup fraction of $1.6$\% at the pulsar location. 
Since the pileup is low, we improve the spatial resolution of the ACIS
image by applying an algorithm to correct the position of split pixel
events \citet{met01}; this decreases the on-axis PSF width in our data
set by $\sim 13$\%.

        In order to determine the background count rate, we located a region of
the S3 chip relatively free of diffuse
emission, cut out point sources, and measured a count rate of
0.59 counts\,s$^{-1}$ in the 0.5-7 keV band. 
Since much of the S3 chip is covered by diffuse nebulae, we split the background
region into two parts: the southern corner of the chip, as well as a
portion of the chip northwest of the pulsar, exterior to the extended emission. 
The nominal background count rate
for the S3 chip in this band is 0.32 counts\,s$^{-1}$. Although no large
background flares were seen, solar activity was high during the observations
which may contribute to the increased background. It is also likely that
low surface-brightness diffuse emission from either extended PWN structure
or Galactic background makes a non-negligible contribution.

\begin{figure}[h!]
\plotone{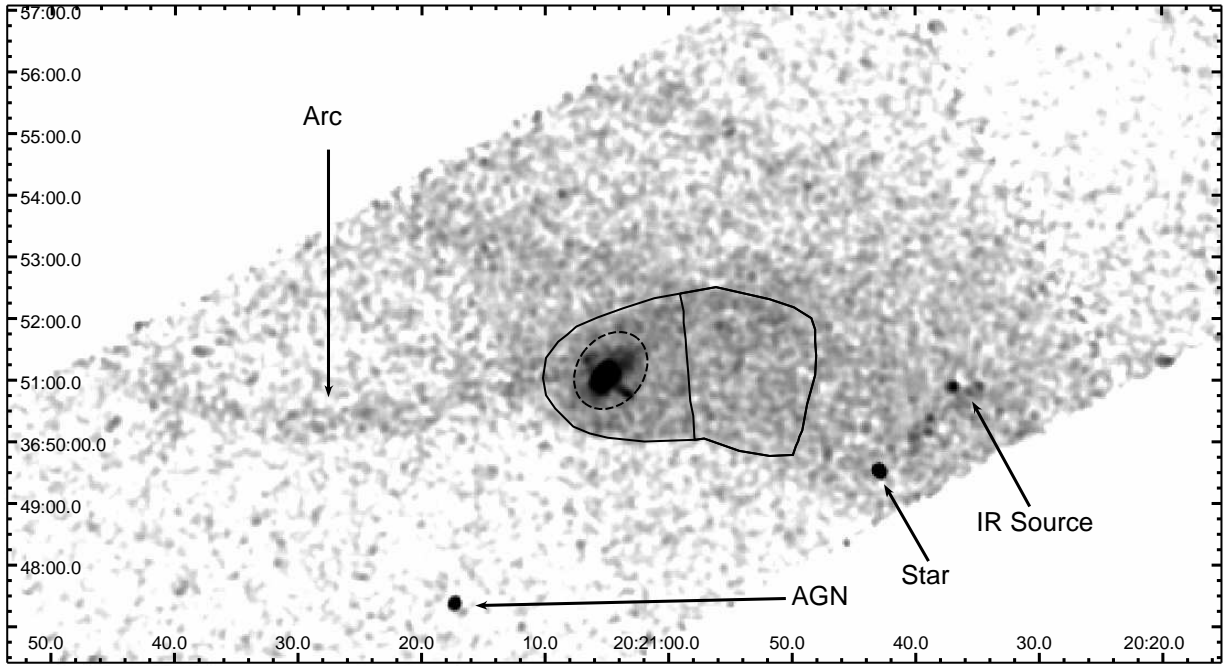}
\caption{
ACIS-S 1-7 keV image (exposure corrected, 
6$^{\prime\prime}$ Gaussian smoothing). Point sources have been
removed except for the AGN candidate, a bright USNO star, and
four sources coincident with a diffuse IR nebula. Extraction regions
for the outer nebula (and dotted exclusion region) are also shown.
}
\end{figure}
               
\subsection{X-ray Spatial Analysis}

        In order to bring out the large scale diffuse emission, we
excised point sources (retaining the pulsar and a few other sources
of interest), 
 exposure-corrected the
image to minimize chip gaps, and smoothed with a 6\arcsec\ FWHM
Gaussian.  Figure 1 reveals considerable emission surrounding the pulsar
located at R.A. = 20:21:05.43, decl. = +36:51:04.63.
A thin arc of emission extends $7\farcm7$ east from the pulsar
to the edge of the S2 chip. To the west, diffuse emission covers
much of the S3 chip and extends onto S4.  The southern
boundary of this emission is very pronounced, with a sharp arc
bracketing the pulsar to the east and south.

The pulsar is surrounded by an axisymmetric inner nebula extending
20$^{\prime\prime}$ $\times$ 10$^{\prime\prime}$ with two ridges
of emission  flanking the pulsar point source.  Southwest of the
pulsar is a very bright inner jet 4$^{\prime\prime}$ long, while
to the northeast a similar, albeit dimmer, feature forms an
inner counter jet. On larger scales the jet can be followed in
a fainter extension from 4$^{\prime\prime}$ to 10$^{\prime\prime}$,
and intensifies to a bright, narrow outer jet running
12$^{\prime\prime}$ to 30$^{\prime\prime}$ from the pulsar.
This outer jet ends in a distinct knot of emission. On the opposite
(`counter-jet') side the extended jet is not obvious. However,
there is a hard, unresolved source 23$^{\prime\prime}$ from the pulsar
roughly along the counter-jet axis with no obvious stellar
counterpart; this is plausibly the counterpart of the jet knot to
the south.  The emission from the outer jets may be variable. For
example \citet{het04} did not conclusively detect the outer jet. In obs0,
the structure is not visually apparent with
9.0$\pm{4.7}\times10^{-4}$ counts\,s$^{-1}$ from this region,
while in the new data the outer jet produces a significant detection
with 1.3$\pm0.2\times10^{-3}$
counts\,s$^{-1}$ above a surrounding background annulus.  There is also
no detection of the jet terminal knot in obs0. In contrast, \citet{het04}
describe a diffuse structure on the counter-jet side. We remeasure this
region and find  1.9$\pm 0.5\times10^{-3}$ counts\,s$^{-1}$ in obs0; the
same region in the new data contributes 1.0$\pm{0.2}\times10^{-3}$
counts\,s$^{-1}$ above the local background. While the excess is formally
significant, there is no obvious coherent structure extending to the
northern knot.  Of course, variation in extended jet structure is not
unexpected: the Vela pulsar outer jet displays dramatic variability
on timescales of days to weeks \citep{pet03}.

\begin{figure}[h!]
\plottwo{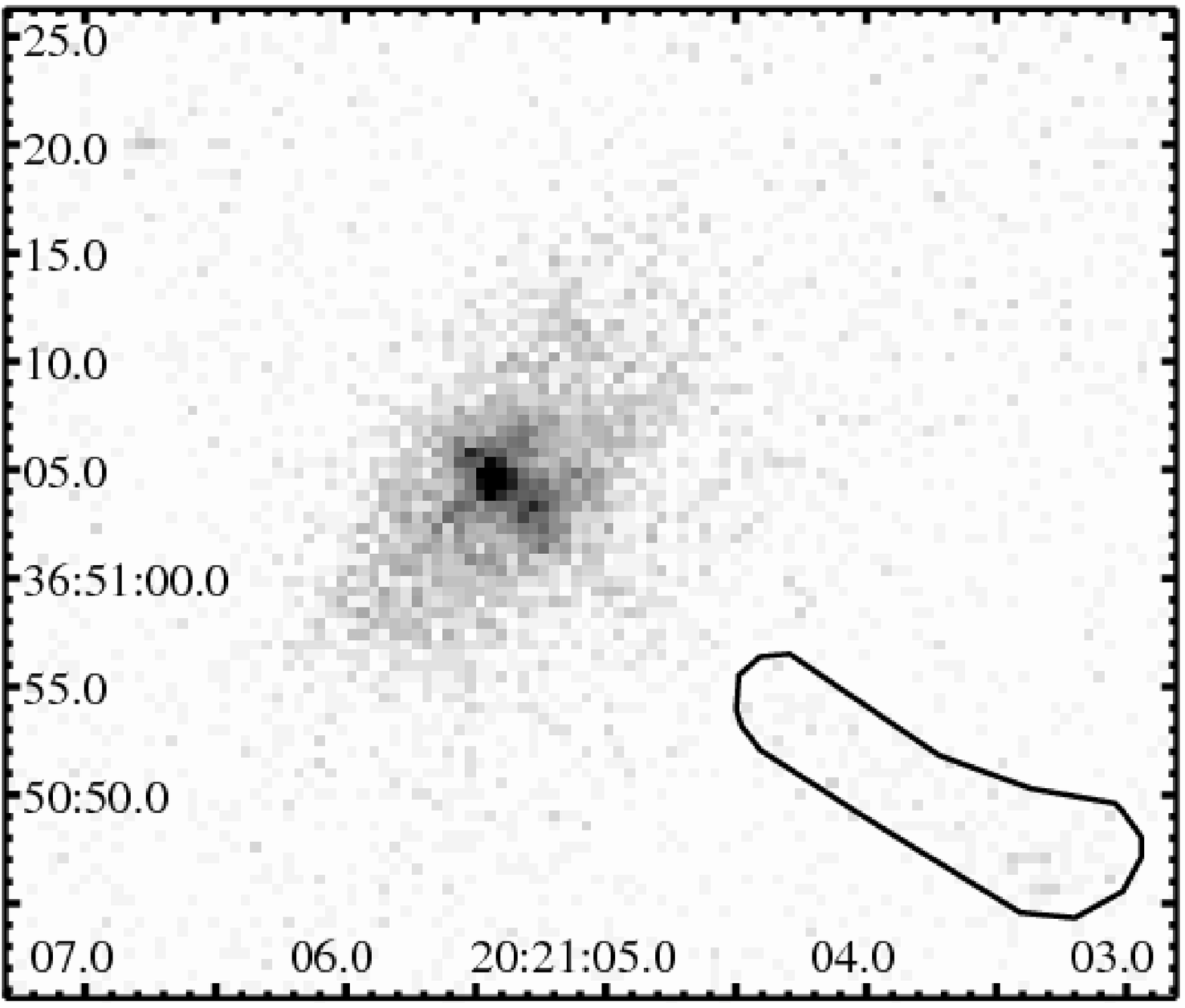}{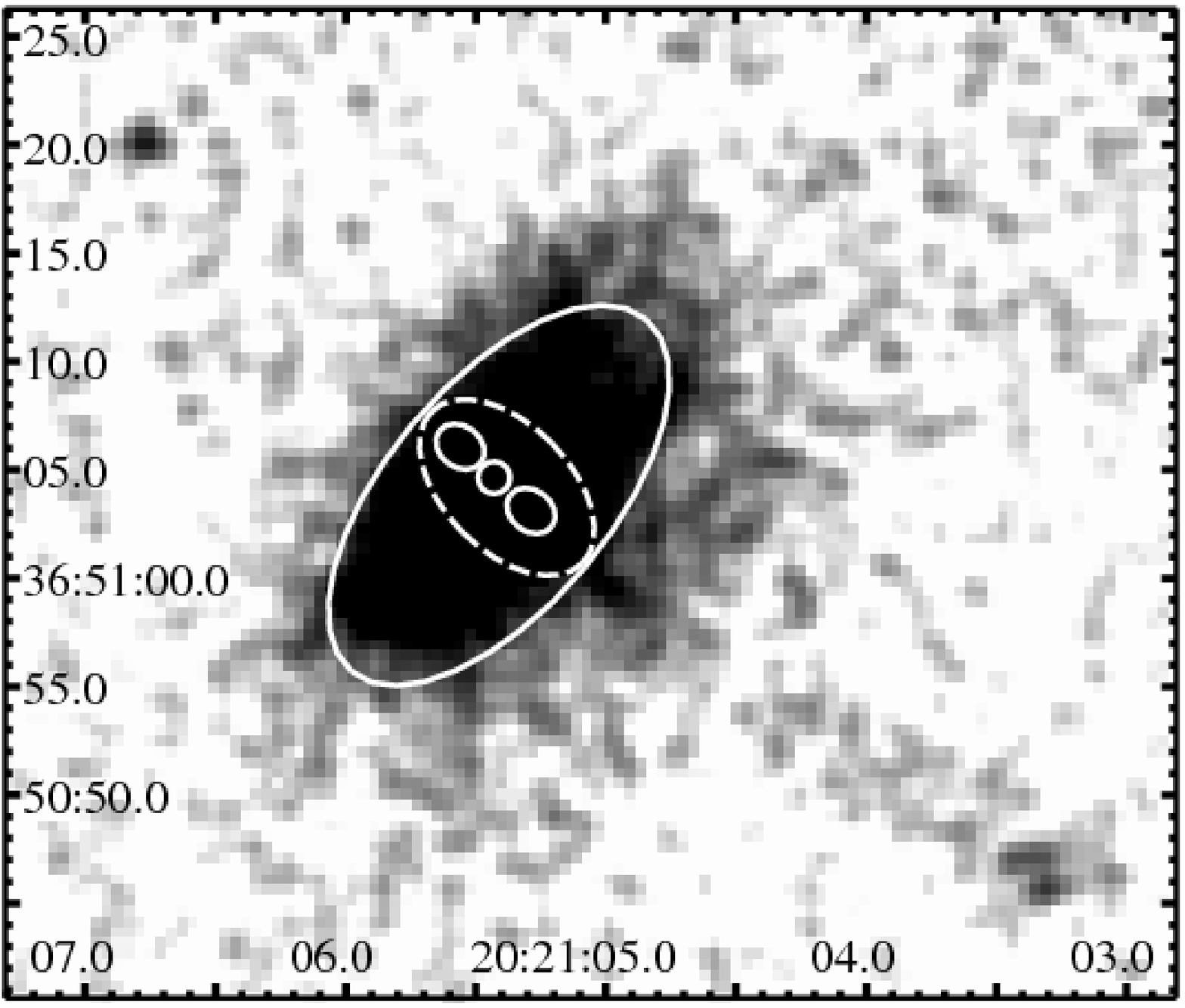}
\caption{
1-7 keV images. Left: A soft stretch shows the double `ridge' structure of the
equatorial PWN, along with the extraction aperture of the outer jet.
Right: A deeper stretch and $1^{\prime\prime}$ Gaussian smoothing bring
out the diffuse emission and jets. The extraction apertures for
the pulsar, inner jets and equatorial PWN (with dashed exclusion region)
are shown against the PWN.
}
\end{figure}

        Turning to the central, brightest portion of the PWN, we see
two ridges of X-ray emission perpendicular to the jet symmetry axis,
one crossing just above the pulsar position and one crossing the bright
inner jet. As noted in \citet{het04}, this suggests a double
torus structure, like that of Vela, but seen more nearly edge-on.
We further explore the PWN structure by fitting torus models in \S2.2 below.
There is a tradition of naming PWNe after animals, dating back to
the fancied resemblance of the remnant of SN1054 in a drawing from
visual observations by William Parsons, Earl of Rosse (1844) to,
variously, a crab or a crab's claw.
For the J2021 PWN, we have a double ridge extending to each side of the
pulsar and a single bright narrow jet. We thus christen this the
`Dragonfly Nebula' with the double ridge forming the paired wings and
the outer jet forming the tail.

As usual in any deep Galactic exposure, several dozen unresolved sources, many
coincident with field stars, are detected in our image. We mention here only the
sources of particular interest. One of the brightest sources in the field,
CXOU J202117.4+364723.7, has a hard, absorbed spectrum and no optical,
IR, or radio counterpart. We consider this a likely AGN viewed through the plane.
Roughly $4\farcm7$ southwest of the pulsar, a strong X-ray source is
seen coincident with the bright field star USNO-B1.0 1268-00448692.

Examination of MSX 8$\mu$m data shows a cavity coincident
with the X-ray nebula. In particular, the southern edge of the nebula
coincides with a fairly sharp step in the 8$\mu$m flux (Figure 3). This
can be interpreted as a dust deficit in the X-ray nebula interior, caused by
evaporation of embedded dust grains. The trail also seems to correlate with a
faint 1.4GHz continuum in the DRAO GPS maps, but more sensitive imaging would
be needed to confirm the reality of this structure.  At the west end of this
cavity the MSX 8$\mu$m data shows a peculiar double-lobed
diffuse ($1\farcm7\times 2\arcmin$) object,
centered at 20:20:36.1 +36:50:37 (Figure 3). The center of this object is within
$40^{\prime\prime}$ of the $S_{1.4GHz} = 25.8\pm4.1$\,mJy source NVSS J202036+365123.
Intriguingly \citet{pet07} noted a double-lobed mid-IR/radio nebula of similar size
near PSR B1823$-$13 and its PWN, although the pulsar proper motion argues against
association in this case. Four of the faint X-ray sources are coincident with
the IR nebula.  Three of the X-ray sources lie within the lobes,
while the fourth is in the bisecting dark lane. All appear absorbed with
virtually no counts below 1 keV; the association with the IR complex,
if any, is not obvious. 

\begin{figure}[h!]
\plottwo{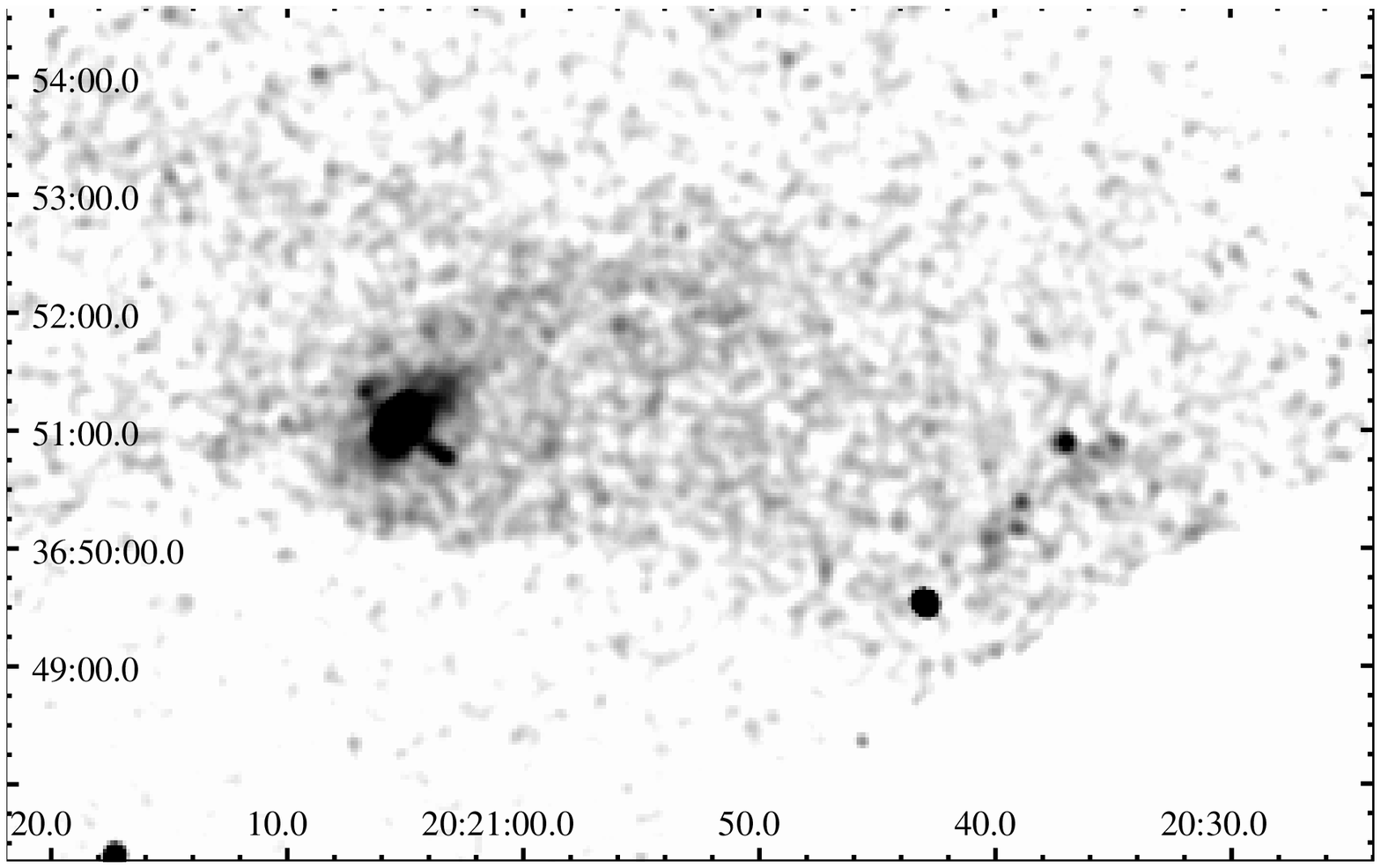}{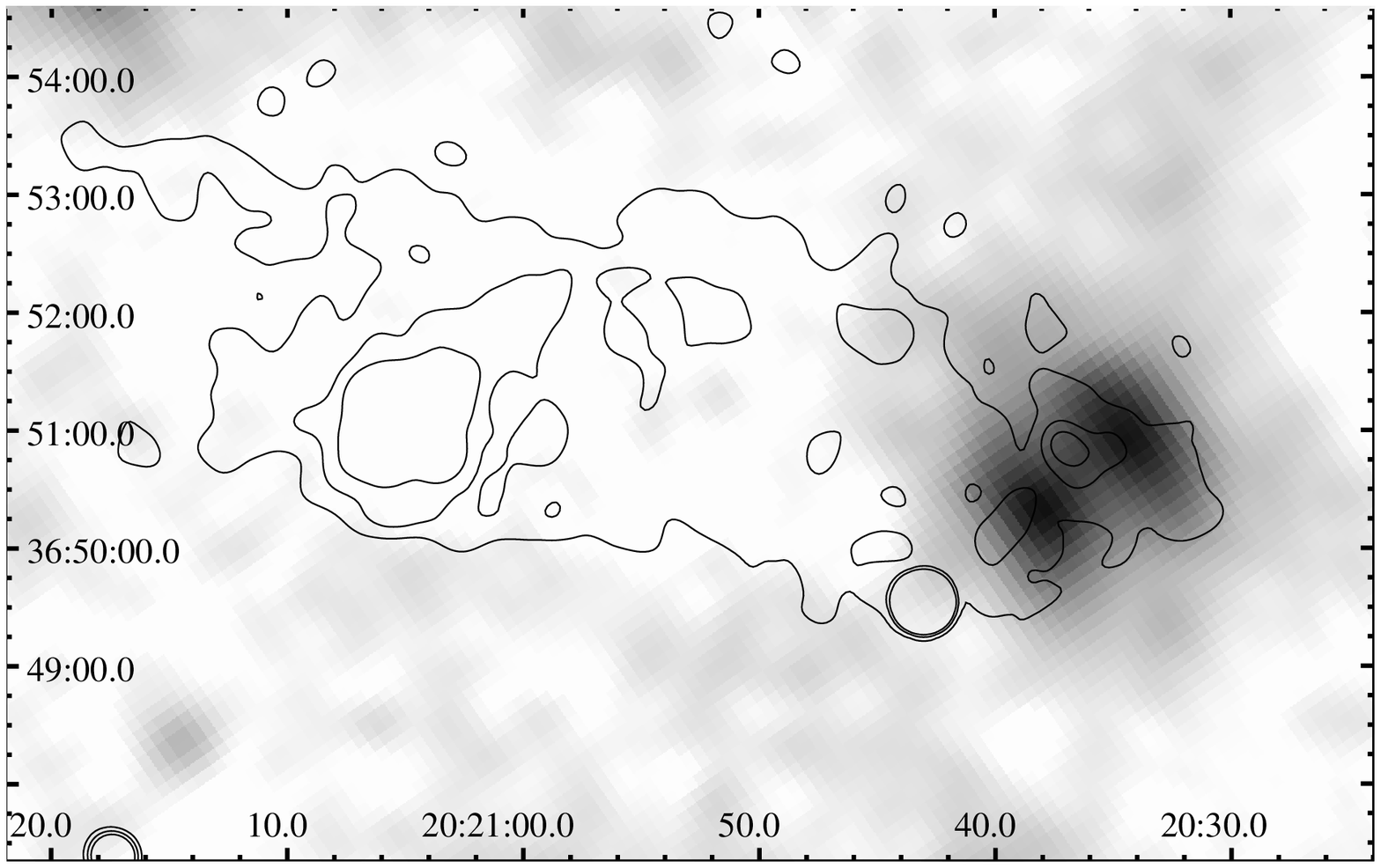}
\caption{
Left: ACIS image with $3^{\prime\prime}$ Gaussian smoothing showing the
PWN, jets, and outer nebula
trailing off to the west. Right: MSX A-band image (8.3\,$\mu$m) overlaid
with heavily smoothed contours from the X-ray data -- a
8.3\,$\mu$m deficit follows the outer PWN; faint X-ray point sources
are found in the bright double-lobed IR source.
}
\end{figure}

\subsection{Spatial Modeling of the Inner Nebula}

In the introduction we noted that many PWNe display axisymmetric torus and
jet structures formed from a relativistic wind with
significant latitudinal variation. MHD simulations of such winds
\citep{kl04,det06,buc07} show that Doppler boosted arcs and jets
can appear in synchrotron emissivity maps, often associated with
lines-of-sight tangential to discontinuities in the flow. For
the brightest PWNe (i.e. Crab and Vela) the {\it CXO} images are
sufficiently detailed for comparison with these models. For more
typical sources, only the overall scale, orientation and Doppler
enhancement can be measured. These parameters can be useful for probing
several aspects of pulsar physics \citep{nr08}.
Even when the PWN structures are less than visually stunning,
\citet{nr04} showed that it is possible to extract robust values for
the basic angles and scales by fitting a simplified Doppler boosted torus
model to the \emph{CXO} images. Following their methodology,
we fitted the `Dragonfly Nebula' with a double torus model similar to
that of the Vela PWN, including a point source and a uniform background.
The fitting minimizes residuals using a Poisson-based likelihood function.
We estimated the statistical errors by re-fitting Monte
Carlo simulations of the best-fit model. The systematic errors are
estimated by in turn re-fitting the data with the jets and point source regions
removed; see \citet{nr08} for details.

For comparison, we also tried fitting with a single torus model;
in this case, a large torus thickness, or `blur', is required for a viable fit.
This single torus model does a substantially poorer job of matching
the inner structure near the pulsar. For example, if we exclude the pulsar
and jets and collapse the counts along the symmetry axis there
is a clear dip between the tori in the nebula brightness profile,
0.5-2\farcs5 SW of the pulsar. In this region the single torus model
departs from the data at the 6$\sigma$ level. However the large
`blur' in the single torus model allows it to fit the general
excess of diffuse emission skirting the PWN better than the sharper
two-torus model (see Figure 4). This allows the fit statistic to be
of comparable quality. Thus with a free fit, the double torus model
is preferred, but not definitively so. If an extra 'halo' component
were added to the model, the preference for the double torus would be
stronger, but this is not really required by the present data quality.

Figure 4
shows the data in 0.5-8\,keV band compared to the
best-fit models, with the parameters listed in Table 1:
the axis position angle $\Psi$, inclination angle $\zeta$, torus radius
$R$, blur $\delta$ for the torus profile, post-shock flow velocity $\beta$,
and separation $d$. The best-fit orientation parameters are consistent in
both models, though $R$ changes slightly due to the different blur.
The new results also show good agreement with the previous fits
by \citet{het04}, but provide more precise measurements.
The relatively small errors indicate that the fits are robust and
not sensitive to the presence of the jets.
In summary, the numerical fits confirm our visual impression that
the double structure better matches the data. However, unless we
allow extra model components to absorb the extended halo counts
or fix the blur $\delta$ at small values for both models, the fit
statistic for the single torus is not dramatically worse.  Note that
the Poisson statistics employed here preclude a direct comparison of
the two models with the $F$-test.

\begin{table}
\begin{tabular*}{\textwidth}{c|cc}
& double tori & single torus \\ \hline
$\Psi\ (\arcdeg)$ & $50.1\pm0.4\pm0.6$ & $50.0\pm1.1\pm0.11$ \\
$\zeta\ (\arcdeg)$ & $85.9\pm0.2\pm1.0$ & $84.9^{+0.4}_{-0.3}\pm0.2$\\
$R \ (\arcsec)$ & $11.1\pm0.2\pm1.0$ & $9.3^{+0.3}_{-0.2}\pm1.0$ \\
blur (\arcsec) & 1.2 (fixed) & $2.9\pm0.04\pm0.2$ \\
$\beta$ & $0.84\pm0.01\pm0.001$ & $0.78\pm0.01\pm0.03$ \\
$d$ (\arcsec) & $3.7\pm0.04\pm0.12$ & - \\
Point Source (cts) & $1240$ & $1209$ \\
Torus (cts) & 3896 & 4459 \\
\end{tabular*}
\label{table:fit}
\caption{Best-fit torus parameters for the double tori and single torus
models, with 1$\sigma$ statistical and systematic errors.}
\end{table}

\begin{figure}[h!]
\includegraphics[angle=-90,scale=0.8]{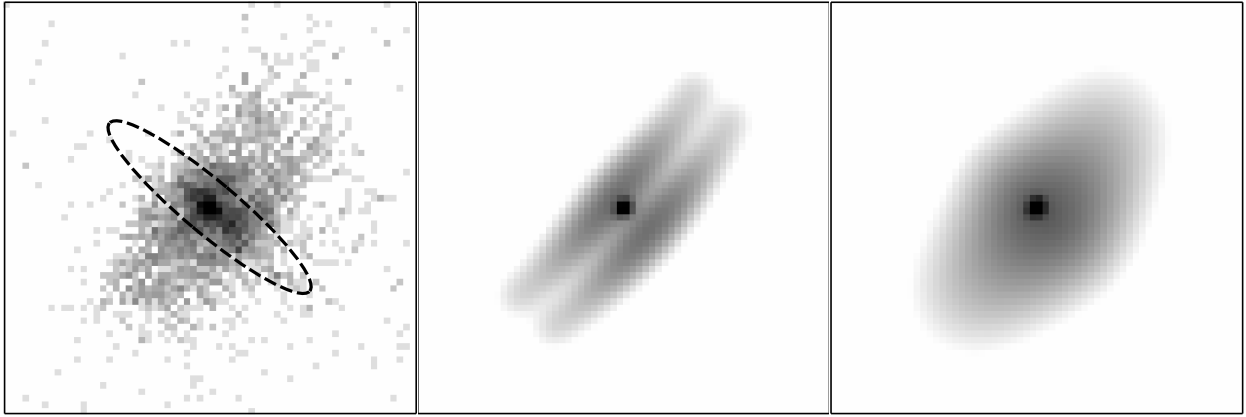}
\caption{
0.5-8\,keV image compared to the best-fit double tori
and single torus models. The dotted ellipse shows the region
excised to estimate systematic errors.
}
\end{figure}

\subsection{Spectral Analysis}

We used CIAO's Sherpa environment (version 3.4) and XSPEC (version 12.3.1) to
fit the spectra of the pulsar and nebula.  For consistency, we bin  all
spectra to a signal-to-noise ratio of three; as a result the jet was
fit only with the new data sets, as the old observation (obs0) has insufficient counts.
The extended emission lies outside the field of view of obs0, and therefore must
also be fit with the new data sets alone. Careful background subtraction is
important for robust spectral fitting; for the inner jets we subtract the surrounding
inner nebula as background, for the star and AGN we utilized surrounding annuli,
and for all other regions we used a large source-free background region.
This background region is on the S3 chip for all spectral fits except for
the eastern wisp, which uses a background region on the S2 chip. 
Fit errors are projected multi-dimensional values at the 90$\%$ confidence level,
except for fluxes. Multi-dimensional flux error estimates are often quite
large due to spectral parameter uncertainties, so we list single parameter
errors at the 90$\%$ confidence level.

We determined the interstellar absorption for the pulsar/PWN
complex by simultaneously fitting the inner and outer components of the nebula with
a global $N_{\rm H}$, but different power-laws; this yields a hydrogen column
density of  $6.7_{-0.7}^{+0.8}$$\times 10^{21}$ $ {\rm cm^{-2}}$, consistent
(albeit with much smaller errors) with
the \citet{rrk01} value of $5.0\pm2.5$$\times 10^{21}$ $ {\rm cm^{-2}}$,
\citet{ret02} value of $N_{\rm H}$ = 7.6$_{-3.5}^{+4.7} \times 10^{21}$ ${\rm cm^{-2}}$
  and the
\citet{het04} value of $7.8_{-1.4}^{+1.7}$$\times 10^{21}$ ${\rm cm^{-2}}$. 
The value of  $6.7\times 10^{21}$ ${\rm cm^{-2}}$,
is adopted in all final fits to components of the PSR/PWN complex.
Although the soft thermal spectrum of the pulsar itself
provides many photons, the effective temperature is highly covariant with $N_{\rm H}$;
accordingly fits to the relatively simple power laws of the extended non-thermal
emission provide the best absorption constraints. The two brightest
point sources have absorptions bracketing the pulsar value. The field
star B1.0 1268-00448692 gave a low $N_{\rm H}$ of 0.2 $\times 10^{21}\,{\rm cm^{-2}}$.
A power-law fit to our candidate AGN gives $N_{\rm H}$ = 2.2 $\times 10^{22}$ ${\rm cm^{-2}}$,
$\sim3\times$ larger than the PSR/PWN value. This supports its identification as
an extragalactic source, although some of the absorption may be intrinsic.
       
\subsubsection{Pulsar Spectrum}

To measure the spectrum of the point source with minimal nebular
contamination, we define a source region of radius 1.5 pixels
(0\farcs74).  Pulsar spectral fitting is complicated by the
fact that the inner nebula surface brightness scales inversely
with radius; extrapolating the radial profile indicates that
the nebular surface brightness is $\sim2.5\times$ greater in
the source region than in an annulus extending 2.5 to 6 pixels
radially from the pulsar.  This precludes simple annular background
subtraction. Instead we determine the nebular power law in this
surrounding annulus and scale up the amplitude as predicted from
the brightness profile, giving an unabsorbed 0.5-8 keV flux of
4.1$\times10^{-14}{\rm erg/cm^2/s}$ and $\Gamma$=1.20.
We hold this amplitude and $\Gamma$ fixed as a
contribution to the background in the pulsar aperture, while fitting
the pulsar point source spectral model.  Both the pulsar and
annulus regions utilized the large source-free background region on
the S3 chip.

For point source fluxing, we modeled the necessary aperture correction
by simulating 10 monochromatic PSFs from 0.5 to 9.5 keV with the
Chandra Ray-Trace program, ChaRT.  The enclosed energy fraction as
a function of radius was then calculated to correct the ARFs used
in the spectral fit.  We fit four different models: a blackbody (BB),
a power law plus blackbody (PL+BB), a magnetized neutron star hydrogen atmosphere (NSA),
and a power law plus neutron star
atmosphere (PL+NSA) \citep{zet96}, all with the extra fixed power law
component from the nebular background. As a check, we also fit
allowing $N_{\rm H}$ to vary. The fitted model parameters were
consistent with those found using the $N_{\rm H}$ fixed by the
nebula fits, albeit with increased errors. For multi-component
models, covariance between components will induce larger effective
flux errors -- e.g. for the power law plus atmosphere model
the power law component flux errors increase from $\sim$22\% to
$\sim$26\% allowing for component covariance (spectral parameters
fixed). The errors grow to $\sim$100\% if all spectral parameters are free.

\begin{table}[!h]
\caption{Spectral Fits to PSR J2021$+$3651}
\tabletypesize=\scriptsize
\begin{tabular}{lcccc|ccccc}
& & Power Law & & & & & BB/NSA \\
Model&$N_{\rm H}$&$\Gamma$&abs.flux&unabs. flux&T$^\infty$&R$^\infty$&abs. flux&unabs. flux&$\chi^2$/dof \\
+PL\tablenotemark{*}&$10^{21}$cm$^{-2}$& &$f_{0.5-8}$\tablenotemark{\dagger}&$f_{0.5-8}$\tablenotemark{\dagger}&MK&d$_{10}$\tablenotemark{\ddag} km&$f_{0.5-8}$\tablenotemark{\dagger}&$f_{0.5-8}$\tablenotemark{\dagger}& \\
\hline

BB & $6.70$\tablenotemark{*} & - & - & - & $2.15_{-0.15}^{+0.16}$ & $4.62_{-0.87}^{+1.40}$& $0.29\pm0.024$ & $1.83\pm0.15$ & 62.7/81 \\

NSA & $6.70$\tablenotemark{*} & - & - & - & $0.88_{-0.02}^{+0.57}$ & $13.1$\tablenotemark{*\Diamond} & $0.28_{-0.030}^{+0.018}$ & $2.95_{-0.32}^{+0.19}$ & 74.9/81 \\

PL+BB& $6.70$\tablenotemark{*} & $1.73_{-1.02}^{+1.15}$ & $0.33\pm0.032$ & $0.48\pm0.048$ & $1.85_{-0.24}^{+0.20}$ & $6.96_{-1.66}^{+4.03}$& $0.23\pm0.023$ & $1.95\pm0.19$ & 32.6/79 \\

PL+BB& $4.33_{-1.75}^{+2.36}$ & $0.99_{-1.40}^{+1.30}$ & $0.32\pm0.093$ & $0.36\pm0.11$ & $2.33_{-0.53}^{+0.56}$ & $2.65_{-2.53}^{+8.74} $& $0.26\pm0.025$ & $0.89\pm0.085$ & 30.1/78 \\

PL+NSA& $6.70$\tablenotemark{*} & $1.92_{-1.81}^{+0.95}$ & $0.32\pm0.069$ & $0.52\pm0.11$ & $0.85_{-0.093}^{+0.25}$ & $13.1$\tablenotemark{*\Diamond}& $0.22_{-0.024}^{+0.022}$ & $2.45_{-0.27}^{+0.24}$ & 32.0/79 \\

PL+NSA& $7.19_{-2.61}^{+1.77}$ & $1.70_{-1.96}^{+0.99}$ & $0.32\pm0.076$ & $0.48\pm0.11$ & $0.87_{-0.24}^{+0.72}$ & $13.1$\tablenotemark{*\Diamond}& $0.24_{-0.024}^{+0.023}$ & $2.85_{-0.28}^{+0.27}$ & 31.2/78 \\

\hline
\end{tabular}
\tablenotetext{*}{held fixed}
\tablenotetext{\dagger}{0.5-8\,keV fluxes in units of $10^{-13}{\rm erg/cm^2/s}$}
\tablenotetext{\ddag}{at a distance of 10\,kpc}
\tablenotetext{\Diamond}{yields a distance of 2.1\,kpc.}
\label{psrspec}
\end{table}

Adding a second thermal component to the simple blackbody model
improves the fit, but since the
pulsar has significant counts above 3keV, a high temperature
$T_2\sim 1.5$\,keV and small area $A_2 \sim 10^{-5}A_1$ would be required.
A similar result is achieved for a two temperature neutron star atmosphere fit.
The small area of the secondary component does not seem compatible with
a polar cap for a Vela-type
pulsar. The power law plus blackbody and power law plus neutron
star atmosphere models also yield much better fits than the
simple absorbed blackbody or NSA models. This point source non-thermal PL
component may represent either magnetospheric emission or unresolved
PWN structure. We favor the former, since the component contributes
more flux than the nebular extrapolation. Although the nebular extrapolation
is uncertain, varying over a plausible range only induced $\sim 10$\%
changes in the point source PL flux.


Our PL+BB model returns a k$T_\infty$ of 0.16 $\pm0.02$ keV which matches well
to the \citet{het04} fit of 0.15 $\pm0.02$ keV.  This value is also comparable to
that of the similarly-aged
Vela pulsar \citep[$\tau_c=11$ kyr, k$T_\infty$=0.128$\pm0.003$ keV,][]{pet01}
and B1706$-$44 \citep[$\tau_c=17$ kyr, k$T_\infty$=0.17$\pm0.02$ keV,][]{ret05}.
The thermal fits allow us to calculate an effective radius as a
function of distance, and hence the emitting area.  For the moment we set
the scale with a nominal distance of 10\,kpc (so $d=10d_{10}$\,kpc),
although as we see below smaller distances are preferred.
Our fit to a power law plus blackbody with $N_{\rm H}$ fixed gives
R = $7.0_{-1.7}^{+4.0}d_{10}$ km,
representing emission from 29$_{-12}^{+42}\%$ of the stellar surface for a
$R_\infty$ = 13.1 km star.  The minimum emitting area of
8.8$\times$10$^{11}$$d_{10}^{2}$ ${\rm cm^2}$ is far greater than
that of the traditional polar cap A$_{pc} = 2{\pi}^2{R}^3/cP
\approx 6.3\times10^{9}$ ${\rm cm^2}$ for any reasonable distance.

        Blackbody deviations for NSA models in principle allow one to determine both
radius and surface redshift, although in CCD quality data these are highly
degenerate.  Consequently, in fitting model atmospheres we fix the surface radius at
$R_s = 10$\,km, corresponding to $R_\infty = R_s (1-2GM/R_sc^2)^{-1/2} = 13.1$\,km. 
The neutron star mass is fixed at M = 1.4 M$_{\odot}$, and we use models with
a surface magnetic field of 10$^{12}$G.  With the mass and radius
fixed, the pulsar distance is determined directly from the model normalization.
With $N_{\rm H}$ fixed our NSA+PL fit yields a distance of 2.1$_{-1.0}^{+2.1}$\,kpc,
where the distance error includes multi-parameter uncertainty.
As usual the NSA model's Wien excess requires a lower $T_{\rm eff}$ and a
smaller distance than that found from blackbody model fits.
For a similar NSA+PL fit 
\citet{pet01} report a $T_\infty$ of 0.059 keV from Chandra ACIS-S
observations of the Vela pulsar.  These authors infer a distance of
220$\pm20$s\,pc from their neutron star normalization,
which is within 25\% of the accepted $\sim300$ pc distance.  Our J2021 neutron
star atmosphere normalization distance estimate of 2.1 kpc therefore lends
some credence to the assertion that the pulsar is significantly closer than 12 kpc.

\subsubsection{Extended Source Spectra}

As a check, we fit the spectra  for the inner and outer nebulae independently,
allowing $N_{\rm H}$ to vary; we find $N_{\rm H} = 6.7_{-1.0}^{+1.1}\times 10^{21}$ ${\rm cm^{-2}}$
and $N_{\rm H} = 6.6_{-1.0}^{+1.1}\times 10^{21}$ ${\rm cm^{-2}}$ respectively,
fully consistent with the global absorption fit above.
Using the fixed global $N_{\rm H}$ allows us to extract useful spectral index estimates
for various components of the PWN complex. The large arc east of the pulsar
may or may not be associated with J2021. Accordingly we also fit this spectrum
allowing $N_{\rm H}$ to vary, though with the larger errors the result is not very
constraining.

\begin{table}[!h]
\caption{Spectral Fits to Other Sources}
\begin{tabular}{lcccccc}
Region & $N_{\rm H}$ & $\Gamma$& abs. flux& unabs. flux & $\chi^2$/dof & obs $\#$ \\
& $10^{21}$cm$^{-2}$& & $f_{0.5-8}$\tablenotemark{\dagger}&$f_{0.5-8}$\tablenotemark{\dagger}& \\
\hline
Inner Neb & $6.7$\tablenotemark{*} & $1.45\pm0.09$ & $3.2\pm0.14$ & $4.2\pm0.19$ & 84.8/200 & 0,1,2 \\
\hline
Inner Jet& $6.7$\tablenotemark{*} & $1.09_{-0.34}^{+0.35}$ & $0.38\pm0.06$ & $0.46\pm0.07$ & 17.0/31 & 0,1,2 \\
Inner Co-Jet& $6.7$\tablenotemark{*} & $1.01\pm0.58$ & $0.23\pm0.05$ & $0.27\pm0.06$ & 10.7/19 & 0,1,2 \\
Jet & $6.7$\tablenotemark{*} & $1.68_{-0.44}^{+0.47}$ & $0.16\pm0.03$ & $0.23\pm0.05$ & 9.1/13 & 1,2 \\
\hline
Outer Neb & $6.7$\tablenotemark{*} & $1.82\pm0.10$ & $7.2\pm0.29$ & $10.9\pm0.45$ & 417.9/655 & 1,2\\
Outer Neb - East & $6.7$\tablenotemark{*} & $1.69\pm0.13$ & $3.9\pm0.23$ & $5.7\pm0.33$ & 207.7/400 & 1,2 \\
Outer Neb - West & $6.7$\tablenotemark{*} & $1.93\pm0.13$ & $3.1\pm0.18$ & $5.0\pm0.28$ & 300.1/477 & 1,2 \\
\hline
Arc & $6.7$\tablenotemark{*} & $1.66_{-0.24}^{+0.25}$ & $0.96\pm0.12$ & $1.4\pm0.17$ & 80.4/151 & 1,2 \\
Arc & $8.8_{-9.9}^{+5.4}$ & $1.87_{-0.48}^{+0.59}$ & $0.92\pm0.11$ & $1.5\pm0.18$ & 79.8/150 & 1,2 \\

\hline
AGN & $25_{-14}^{+28}$ & $1.84_{-1.24}^{+1.43}$ & $0.44\pm0.07$ & $0.94\pm0.15$ & 9.4/17 & 1,2 \\
AGN & $22_{-8}^{+10}$ & $1.7$\tablenotemark{*} & $0.45\pm0.07$ & $0.85\pm0.14$ & 9.5/18 & 1,2 \\
\hline
Star& $0.20_{-0.20}^{+0.10}$ & $3.45_{-0.47}^{+0.62} / 16.1_{-1.5}^{+5.8}$\tablenotemark{\ddagger} & $1.34\pm0.17$ & $1.43\pm0.19$ & 37.1/51 & 0 \\
Star & $0.20_{-0.20}^{+0.10}$ & $3.81_{-0.36}^{+0.58} / 15.8_{-1.2}^{+4.5}$\tablenotemark{\ddagger} & $1.23\pm0.13$ & $1.31\pm0.14$ & 57.2/68 & 1 \\
Star & $0.20_{-0.20}^{+0.10}$ & $2.80_{-0.40}^{+0.43} / 10.1_{-0.82}^{+1.9}$\tablenotemark{\ddagger} & $0.70\pm0.08$ & $0.76\pm0.08$ & 57.3/63 & 2 \\

\hline

\end{tabular}

\tablenotetext{*}{held fixed}
\tablenotetext{\ddagger}{T(MK) for two Mekal thermal components}
\tablenotetext{\dagger}{0.5-8\,keV fluxes in units of $10^{-13}{\rm erg\,cm^{-2}\,s^{-1}}$}
\label{pwnspec}
\end{table}

As one moves progressively further from the pulsar the extended emission exhibits
significant softening, as expected from synchrotron burn-off in the outer PWN. 
The inner jets have the hardest spectra ($\Gamma$=1.1), while the outer jet
region is softer with $\Gamma$=1.7.  The inner nebula  or
`equatorial PWN,' has  $\Gamma=1.5$.  The diffuse trail close to the pulsar
($\approx$30$^{\prime\prime}$-60$^{\prime\prime}$) has a power-law index
of 1.7, while in the farther reaches of the outer nebula we observe an index of 1.9. 

\subsubsection{USNO Star Spectrum}

We estimated a photometric parallax for USNO-B1.0 1268-00448692 using data
from the USNO-B ($\sim$B,R,I bands) and 2MASS catalogs (J,H,K$_s$ bands). 
We converted the survey magnitudes to standard colors and fit the colors to
stellar values \citep{kah95} subject to interstellar reddening \citep{sch98}.
The best global fit to main sequence colors implied class M2 V, with
$A_V \approx 0.45$ at a distance of $\sim 60$pc. The POSS-II `I' band point
has the largest departure from the model colors; if it is excluded the best class shifts
to K6 V with $A_V \approx 1.7$ at $\sim 100$pc. Few post-MS stars of this
color emit coronal X-rays. The best match (excluding I) was a rather implausible
G5 III with $A_V \approx 2.7$ at 1.5\,kpc.

        The star shows modest variability within and between epochs.
We therefore follow the procedure of \citet{get05} and adopt a two
component Mekal plasma model; we fit the
temperatures and amplitudes separately for each observation,
while fitting for a common
$N_{\rm H} = 0.2 \times 10^{21}{\rm cm^{-2}}$.
As expected the star temperatures are higher during the X-ray bright
epochs. The fit $N_{\rm H}$ can be compared with the optical
extinction, where $N_{\rm H}= A_V 1.9\times 10^{21}{\rm cm^{-2}}$ predicts
$A_V= 0.4_{-0.4}^{+0.2}$. The best global interpretation thus appears to be
a relatively mundane nearby M2 V field star; the estimated $N_{\rm H}$ and
$A_V$ are in reasonable agreement, albeit somewhat higher than one would
expect for a 60\,pc distance. However this extinction adds no significant
contribution to the full column to J2021.

\section {Discussion}

\subsection {Distance}

As noted in the introduction, the dispersion measure of 369\,${\rm cm^{-3}}$pc
($N_e$=1.13$\times10^{21} {\rm cm^{-2}}$) for J2021 corresponds to a
distance of 12\,kpc in the \citet{cl02} model. Such a large distance is
problematic on several grounds, so it is important to check this estimate
against other observables.  Our best fit hydrogen column density is
$N_{\rm H}$=6.7$\times10^{21} {\rm cm^{-2}}$, while the total Galactic $N_{\rm H}$
in the direction of J2021 is estimated from H{\sc I} maps to be
1.2$\times10^{22} {\rm cm^{-2}}$ according to \citet{dl90} or 
9.7$\times10^{21} {\rm cm^{-2}}$ according to \citet{ket05}.  Thus we
already detect somewhat less than the expected column, suggesting that J2021
does not reside in the far reaches of the Galaxy. Moreover, with
$N_{\rm H} = 2.2 \pm 0.9\times10^{22} {\rm cm^{-2}}$ measured for our AGN
candidate an even higher column is implicated (although, again, some
of this absorption may be intrinsic). On the balance, these data suggest that
the pulsar/PWN absorption, while large, is $\sim 1/3 - 2/3$ the full
Galactic column density in this direction, placing the source well
inside the Milky way disk.

        According to \citet{cl02}, DM distance estimates are
seldom off by $>50\%$, unless the line of sight intersects anomalously
dense electron clouds. In any case, at such large DM it is difficult for
any one H{\sc II} region to substantially increase the value. Nevertheless,
we note that our observed neutral to ionized ratio $N_{\rm H}$/$N_e \approx 6$
is somewhat below the canonical factor of 10, while \citet{het04}
noted that J2021's radio pulse profile showed anomalous scattering with
a scattering measure SM $\sim 100\times$ that predicted by the NE2001 model.
These provide evidence for high ionization along the line of sight.
However, none of the X-ray structures that we see provide an obvious
source of such ionized gas: the `Arc' passes north of the pulsar and the
extended nebula is too faint to represent a large emission measure
or column density.

One problem with the large 12\,kpc distance would be the high $\gamma$-ray
efficiency.  For an assumed 1sr beaming, the observed 3EG point source
$\gamma$-ray flux gives a 0.1-5\,GeV efficiency of $\eta_{\gamma} \equiv
L_{\gamma}/{\dot E} \sim 0.15d_{10}^2$.  We should compare this with
the similar $\gamma$-ray pulsars Vela (d$\approx300 {\rm pc}$, ${\dot E}$
= 6.9$\times10^{36}$ ergs\,s$^{-1}$) and PSR B1706$-$44 (d$\sim$3 kpc,
${\dot E}$ = 3.4$\times10^{36}$ ergs\,s$^{-1}$) which have 1sr efficiencies of
$\eta = 6.6 \times 10^{-4}$ and $1.8 \times 10^{-2}$, respectively, for the
same energy range and assumed spectral index. If the efficiency matches that of
PSR B1706$-$44 with nearly identical spin properties, we infer a
distance of $\sim 4$kpc. At the very low efficiency inferred for the Vela
pulsar, the source would need to be at a rather implausible 800\,pc.

        We can also use trends in the X-ray luminosities of other
pulsars and PWNe to obtain rough distance estimates. For 6 other Vela-like pulsars
\citet{kpg07} find an efficiency for conversion of spindown power
ranging from $\eta_X \sim 3\times10^{-5}$ to $3\times10^{-4}$ in the 0.5-8\,
keV band. Some of this
spread likely stems from variation in the ambient medium as well as
differing pulsar properties. If we adopt
only the unabsorbed flux of the equatorial PWN and the jets, we
see that the observed $5 \times 10^{-13}{\rm ergs\,cm^{-2}\,s^{-1}}$ 
implies a distance range of 1.3-4.1\,kpc. Inclusion of emission
from the extended outer nebula
does not seem consistent with the flux estimates for other PWNe,
but would shrink the implied distance range to 0.7-2.3 \,kpc.
The non-thermal luminosity of the point source should also
scale with spin-down power; \citet{kpg07} note that these quantities
correlate surprisingly well, with
L$_{PWN}\approx5L_{psr}^{nonth}$. For J2021, the power law flux
is relatively poorly determined, but at only
$0.5 \times 10^{-13}{\rm ergs\,cm^{-2}\,s^{-1}} $
even the equatorial nebula represents $\sim 8\times$ this
component, making it easiest to interpret the equatorial nebula as the full PWN flux
and supporting the 1.3-4.1\,kpc distance.  We caution that this estimate should not
be taken too seriously given the large scatter in efficiency ratios.   

        A final spectral estimate of the distance comes from the
neutron star thermal flux, which if interpreted as magnetic H
atmosphere emission from the full surface of a canonical neutron star,
implies $d \sim 2.1 R_{10}$kpc, for a 10\,km local neutron star radius.
If the surface spectrum is a blackbody plus power law, then larger distances
(4-8\,kpc) are nominally allowed. However the required $\sim2$MK
temperature would be anomalously high for full surface emission of
a Vela-aged pulsar. A polar cap could be this hot, but an implausibly
small $\sim 0.8$\,kpc distance would be required to match the flux
to the canonical polar cap area.  In sum,
though, the spectrum and flux estimates for the X-ray emission from
this pulsar and it's PWN, along with the comparison of the inferred
$\gamma$-ray efficiency with other Vela-like pulsars implies a
distance of 3-4\,kpc, substantially smaller than the DM-inferred
distance. We adopt an estimate of $d=4d_4\,$kpc , with $d_4 \approx 1$
in what follows. 
The source of the extra dispersion and scattering along the line-of-sight
to J2021 is yet to be identified.

\subsection{Birthsite and Environment}

        Most pulsars show bow shock PWNe when traveling at
supersonic speeds through the interstellar medium. However, with
a characteristic age $\tau_c = 1.7 \times 10^4$yr, J2021 should
still reside in its parent supernova remnant, where in the absence of
a central pulsar or PWN the reverse shock reaches the center on a timescale of
$$
t_{Sed} \approx 7 (M_{ej}/10 M_{\odot})^{5/6}
(E_{SN}/10^{51} ergs)^{-1/2} (n_{ISM}/1 {\rm cm^{-3}})^{-1/3}{\rm kyr}
\eqno (1)
$$
\citep{rc84}. In the presence of a young pulsar, the reverse shock
will collide with the expanding PWN in a time somewhat less than this.
If the remnant is just entering the Sedov phase
the reverse shock should be reaching the remnant center,
where it would crush the PWN. For typical high pulsar velocities
the system can be well off-center in the parent
SNR; the crushed PWN should then trail back toward the supernova site,
with the pulsar at the leading edge \citep{vdsdk04}.  Instead,
we see symmetric equatorial structure, with jets extending to
$\sim 30''$ and no prominent bow shock on smaller scales. This
suggests a low velocity pulsar in a relatively quiescent
surrounding medium. It is possible that the reverse shock has not
yet reached the PWN or, alternatively, that the reverse
shock reverberations
settled down some time ago, leaving a relatively uniform high-pressure
SNR interior as the ambient environment.  Given the characteristic age of
the pulsar (17 kyr), and the timescale for the SNR reverse shock to reach the
PWN ($\sim5$ kyr), we favor the latter interpretation. 

        If a $10^{51}E_{51}$erg supernova has produced a Sedov-phase
remnant  of age $10^4t_4$y in an external medium of density $n_{ISM}\,{\rm cm^{-3}}$,
then we expect a shell of radius
$$
\theta_{\rm SNR} \approx 12^\prime (E_{51}/n_{ISM})^{1/5} t_4^{2/5}/d_{4}.
\eqno(2)
$$
While there is no cataloged remnant and no obvious X-ray or radio
shell in the vicinity, we could plausibly associate the thin
arc with a partial SNR shell. However, with a radius of curvature
$\sim 7^\prime$ centered $\sim 8^\prime$ northeast of the pulsar,
it seems unlikely that this represents the limb of the parent
SNR. Of course, the scale would be acceptable for the large $\sim
10$\,kpc distance, but then the pulsar would have left the shell
at 2300$d_{10}$/$t_4$\,km\,s$^{-1}$; the PWN should then show a bow shock
with an unresolved stand-off. It is possible that the arc is a filament
on the face of a larger remnant, with the pulsar close to the center.
A search for a large $\sim 30'$ low surface brightness shell could be
productive.

        Assuming for the moment that the PWN does live in the
heated interior of a Sedov-phase SNR, we expect an ambient pressure
$P_{\rm SNR} \propto \rho {\dot R}^2 = \chi \rho_{ISM}^{3/5} E_{SN}^{2/5}
t^{-6/5}$.  We take $\chi$ = 0.047 \citep{vds00} so that
$$
P_{\rm SNR} \approx 1\times10^{-9} E_{51}^{2/5} n_{ISM}^{3/5}
t_4^{-6/5} {\rm g\,cm^{-1}\,s^{-2}}
\eqno(3)
$$
The PWN equatorial torus represents the termination shock where
the pulsar wind momentum flux balances this pressure. Two factors
complicate our estimation of the termination shock radius for
J2021. First the wind is expected to be equatorially confined so
that the momentum balance condition is
$$
P_{\rm SNR} \approx {\dot E} \xi /(4\pi R_{WS}^2 c)
\eqno (4)
$$
with $\xi >1$. For example, for a wind momentum scaling with latitude
$\theta$ as sin$^2 \theta$ \citep{bk02} one expects
$\xi = 3/2$. Second, while in some PWNe such as the Crab, we observe
the `sub-luminous zone' of unshocked pulsar wind and hence directly
resolve $R_{WS}$, here we only see the overall extent of the equatorial
flow, which is generally 2-3$\times$ larger (e.g. Ng \& Romani 2008).
Accordingly, using the pressure (3) we estimate the overall angular
scale of the equatorial torus as
$$
\theta_{\rm eq} \approx (2-3 \times) 1\farcs6 E_{51}^{-1/5} n_{ISM}^{-3/10} t_4^{3/5}
\xi^{1/2} d_{4}^{-1}.
\eqno (5)
$$

Taking $E_{51}=n_{ISM}=1$, $\xi \approx 2$ and $t_4=1.7$ (the spindown age) we get an
overall scale of $\theta_{eq} \sim (2-3\times)3.1^{\prime\prime} d_{4}^{-1}$.
If we identify the double ridge of the `wings' of the dragonfly with this
overall torus, we have a measured size of $\approx 10^{\prime\prime}$,
which from (5) implies a distance of 2.5-3.7\,kpc.  This is in good accord with
our previous estimate of 3-4 kpc.

        More heuristically, we can simply scale to the observed angular radius
of the overall toroidal flow for Vela ($\sim 30''$) and PSR B1706$-44$
($\sim 15''$). Matching the $\sim 10''$ of the dragonfly wings suggests
$\sim 1$ or $4.5$\,kpc (for Vela and B1706, respectively). Again, while not
precise, the geometrical estimate of the wind shock scale suggests a distance
$\leq 4$\,kpc.

\subsection {Outer Nebula and Jets}

        In Figure 4, we see that the diffuse emission surrounding the
equatorial PWN is brightest in a $\sim 25''$ radius halo around the pulsar,
continuing to the `outer nebula' which forms a  $\sim 50''$ radius
limb-brightened structure surrounding
the pulsar and trailing off to the west, with some curvature. If we interpret
this outer nebula as a bow shock, then with a PWN speed of 10$v_6$\,km\,s$^{-1}$
($v=v_6\times10^6$\,cm\,s$^{-1}$) relative to an
ambient medium of number density $n$, we have
$$
\theta_{BS} \approx [{\dot E} /(4\pi c\rho v_{PSR}^2d^2)]^{1/2} \approx 40'' n^{-1/2}/(v_6d_4)
\eqno (6)
$$
and the observed $\theta_{bs} \sim 50''$ standoff distance implies a rather
low speed of $v_{PSR} \approx 8 n^{-1/2}d_4^{-1}$\, km\,s$^{-1}$. This nebula seems to
extend $\sim 4'$ to the west. If this traces the actual motion since pulsar
birth we require a fairly typical pulsar speed of $v=4'd/\tau_c = 270 d_4$\,km\,s$^{-1}$
assuming the true age is the spindown age. To preserve the large nebula standoff,
the ambient medium should then have a density $n < 10^{-3} {\rm cm^{-3}}$, which
is plausible for the hot interior of a SNR. Some support for this
interpretation comes from the softening of the X-ray spectrum in the western
half of the nebula, which could be attributed to synchrotron cooling of relic particles
at the pulsar birth site.
Alternatively, this outer nebula could be caused by PWNe electrons filling
a pre-existing structure in the SNR interior, e.g. density variations in
a progenitor wind, so it might not trace the pulsar motion.

        Turning to the jet components, it is natural to attribute the flux ratio
between the jet and counter jet to Doppler boosting. For a continuous co-linear
inner jet/counter-jet we expect
$I_j/I_{cj} = [(1-\beta {\rm cos}\zeta)/(1+\beta {\rm cos}\zeta)]^{-\Gamma-1}$,
where $\zeta$ is the inclination of the jet axis to the line of sight.
The spectral fits give $\Gamma \approx 1.1$ and a flux ratio of
$1.7_{-0.5}^{+0.8}$. However, the spatial fit to the
the equatorial structure places it nearly edge-on ($\zeta \approx 85^\circ$),
which gives a maximum flux ratio of $\sim 1.4$ for large $\beta$. Of course
even a small jet misalignment relaxes this stricture; for $\zeta \approx 80^\circ$,
we can accommodate a flux ratio of 1.7 with $\beta=0.7$. Still, with
the jets nearly in the plane of the sky large $\beta$ are required for significant
Doppler boosting.

        A similar argument applies to the outer jets. For the jet an aperture
that avoids the bright knot at the end provides $73\pm16$ counts, while
the same size aperture for the counter-jet has $28\pm14$ counts (not a significant
detection). This leads to a fairly large flux ratio of $2.6_{-0.9}^{+3.8}$.
Of course this cannot be accommodated for an inclination as large as $85^\circ$.
However, visually the jet and counter-jet do appear to be mis-aligned on the
plane of the sky, so a larger counter-jet inclination, allowing larger Doppler
de-boosting, does seem plausible. For example, for the observed jet spectrum
($\Gamma\approx 1.7$) any angle $<75^\circ$ can be accommodated for $\beta< 0.7$.

        In view of the possible jet variability noted above, none of these
flux ratio constraints should be taken too seriously. However,
if we do attribute the faintness of the outer counter-jet to Doppler
boosting, then the presence of a terminal knot for both the jet and counter-jet
could be due to a termination shock, resulting in slower flow with more
isotropic emission. The outer nebula is somewhat brighter in a
$\sim 25''$ region around the PWN and perhaps this
marks the flow boundary where the jets terminate. This would be reasonable
if they lie close to the plane of the sky. In any case the termination probably
lies well within the envelope of the `outer nebula;' with the jets only half the
length of the outer nebula radius, an implausibly small inclination of
$30^\circ$ would be required to make their projected length as small as observed.

\subsection {Energetics and Spectral Results}

        The observed spectral softening from the inner to the outer nebula
(Figure 5) suggests aging of the synchrotron population on the flow time.
At radio frequencies, however, the dominant population
should be uncooled with a spectral index $\alpha_R \approx 0.3$. An
examination of the 1.4GHz DRAO Galactic Plane Survey continuum maps \citep{enget98}
shows a 3$\pm0.5$\,mJy source at the position of J2021, not clearly
resolved at the $\sim$arcmin GPS resolution. Since the pulsar
itself provides only 0.1\,mJy at 1.4\,GHz, we interpret this as PWN emission.

\begin{figure}[h!]
\epsscale{0.6}
\plotone{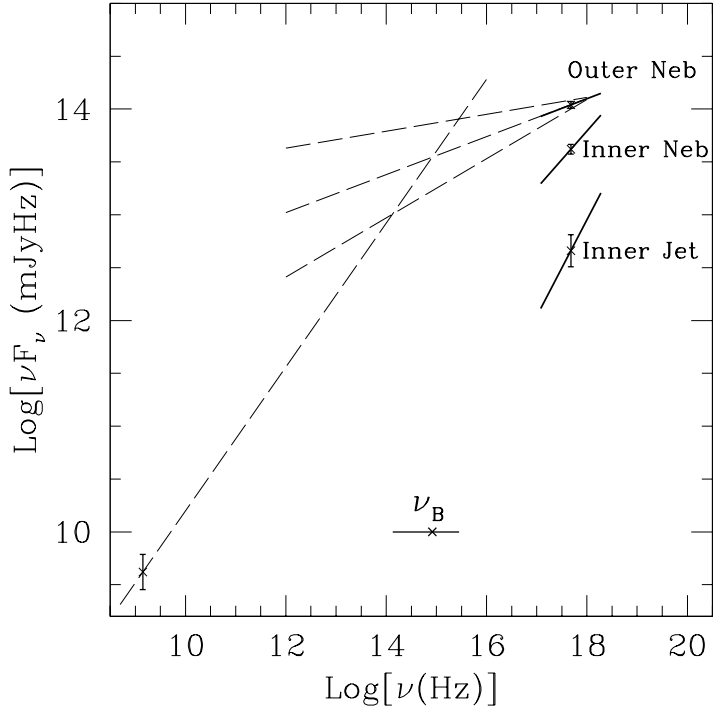}
\caption{
PWN synchrotron cooling. Here we show the radial softening of the
PWN X-ray spectrum (see Table 2). For the dominant flux of the outer
nebula we extrapolate to the observed radio flux with a
$\Delta \alpha$=0.5 cooling break.
}
\end{figure}

        For such a low radio flux, the cooling break must be at relatively
high frequency. Extrapolation of the X-ray flux with a $\Delta \alpha$ = 0.5
break implies Log[$\nu_B$(Hz)] = 14.9$_{-0.8}^{+0.5}$.
The synchrotron break frequency at an age of $10^4t_4$\,yr in a field of $B_{mG}$\,mG 
is $\nu_B = 1.0\times 10^{10}B_{mG}^{-3}t_4^{-2}$\,Hz. Thus we infer an outer nebula
field of $B = 13_{-4}^{+13} \mu$G
for $t_4=1.7$.  This can be compared with the field
extrapolated from $B_s =3.2\times 10^{12}$G at the pulsar surface ($R_s$=10km).
If the field is dipolar within the magnetosphere, followed by $\propto 1/r$
in the wind exterior to the light cylinder ($R_{LC} \equiv c/\omega$)
we expect
$B_{ws}\approx 4 B_s R_s^3/(R_{LC}^2 R_{ws})= 220 \theta_{5}^{-1}d_4^{-1}\mu$G
in the wind termination shock at an angle of 5$\theta_{5}^{\prime\prime}$.
If the field continues to decrease as $\propto 1/r$ in the post-shock flow,
then we would expect a value $\sim 20\mu$G at the $\sim 50\arcsec$ limb of the
outer nebula.  This is in reasonable accord with the photon-weighted
value estimated from the (rather uncertain) spectral break. The corresponding
magnetic pressure $B^2/{8\pi} \sim 2 \times 10^{-11} {\rm g\,cm^{-2}s^{-2}}$ is
somewhat lower than equipartition for a SNR interior. More detailed radio
measurements, resolving the PWN and determining the radio spectral index
would substantially tighten these arguments. It may also be useful to mount
deep optical/IR imaging to detect the PWN flux near the cooling break.

\section {Conclusions}

We observed PSR J2021+3651 and its surrounding structured nebula and
jets with the {\it CXO} ACIS. This `Dragonfly Nebula' displays an
axisymmetric morphology, with bright inner jets, an apparently double-ridged
equatorial inner nebula, and a $\sim 30''$ long polar jet. Surrounding
the central nebula is a low surface-brightness outer nebula. This structure
brackets the pulsar at a radius of $\sim 50''$ and trails off to the
west over 3-4\arcmin. The overall structure is highly reminiscent of
{\it CXO}/ACIS image of the PWN surrounding PSR B1706--44 \citep{ret05},
and, as for this source, is plausibly caused by a low-velocity pulsar
in a relatively quiescent medium. No clear evidence of a host SNR
is seen. IR data does appear to show a dust deficit associated with
the trailed X-ray nebula. 

        Although the photon statistics are limited,
double structure for the equatorial PWN is suggested by both the
initial 19\,ks exposure and, independently, by our new deeper image.
So far Vela is the only other pulsar showing a double ridge
in its PWN. Whether this represents a physical doubling of the equatorial
torus or projection (caustic) effects in an optically thin, Doppler
boosted pulsar wind is still unclear.

        The new spectral data combined with the spatial scale of the PWN
shock and its surrounding nebula do not support the $\sim12$\,kpc
distance suggested by the large pulsar DM. The X-ray absorption,
efficiency arguments, and comparison with other Vela-type PWNe,
while individually inconclusive, all suggest a smaller
$d\sim 3-4$\,kpc. On the balance we prefer this distance scale,
although we find no obvious source of ionized plasma that could
explain an anomalously large DM, so this large value would remain
unexplained.

        This new distance estimate will likely be helpful in interpreting
new $\gamma$-ray observations. At this distance, the GeV
efficiency needed to produce the {\it EGRET} source is comparable to that for
PSR B1706$-$44. Upcoming {\it AGILE}, and especially {\it GLAST}, observations
should test this association. A high quality pulse profile will also be very
useful, as the pulse shape is sensitive to the inclination angle
$\zeta$ estimated here from the PWN fit. For example, in outer magnetosphere
gap models \citep[e.g.][]{ry95}, the large inclination angle of this pulsar
predicts a double peaked $\gamma$-ray light curve with large phase separation.
At higher TeV energies, the PWN may also be detectable, although the striking
morphological similarity to the PWN of B1706$-$44 inspires a note of
caution, since H.E.S.S. searches for B1706$-$44's PWN find only upper limits
of 0.01$\times$Crab \citep{aet05}.
However, it may be relevant that
Milagro finds a source MGRO J2019+37 $\sim 20\pm25'$ west of
PSR J2021+3651 \citep{aet07}. Intriguingly, the offset from J2021
is along the direction of the `outer nebula' trail. Since the
TeV ICS emission is dominated by lower energy, and hence older,
electrons than those that produce the X-ray synchrotron emission,
such offsets are commonly seen for PWNe. A high quality IACT
image from {\it VERITAS} or {\it MAGIC} detecting the PWN
and showing its arcminute-scale structure would be of particular interest.
Conversely, strong upper limits on associated TeV emission would
suggest that some common attribute makes the Compton component of
the `Dragonfly Nebula' and PSR B1706$-$44's PWN anomalously faint.

\acknowledgements
We thank N. Bucciantini for interesting discussions of PWN structure. This work
was supported in part by NASA grant NAG5-13344 and by Chandra
grant GO7-8057 issued by the Chandra X-Ray Center, which is
operated by the Smithsonian Astrophysical Observatory for and
on behalf of the National Aeronautics Space Administration under
contract NAS8-03060.

{\it Facilities:} \facility{CXO (ACIS)}


\begin{thebibliography}{}


\bibitem[Abdo et al.(2007)]{aet07}Abdo, A. A., et al. 2007, ApJ, 658, L33

\bibitem[Aharonian et al.(2005)]{aet05}Aharonian, F., et al. 2007, AA, 432, L9.

\bibitem[Bucciantini(2007)]{buc07}Bucciantini, N. 2007, arXive/0710.0397

\bibitem[Bogovalov \& Khangoulyan(2002)]{bk02} Bogovalov,
S.~V., \& Khangoulyan, D.~V.\ 2002, Astronomy Letters, 28, 373

\bibitem[Cordes \& Lazio(2002)]{cl02}Cordes, J.M. \& Lazio, T.J.W. 2002, astro-ph/0207156

\bibitem[Del Zanna et al.(2006)]{det06}Del Zanna, L. at al. 2006, AA, 453, 621

\bibitem[Dickey \& Lockman(1990)]{dl90}Dickey, J.M. \& Lockman, F.J. 1990, ARAA, 28, 215

\bibitem[Dodson et al.(2003)]{det03} Dodson, R., Legge, D.,
Reynolds, J.~E., \& McCulloch, P.~M.\ 2003, \apj, 596, 1137
 
\bibitem[English et al.(1998)]{enget98}English, J. at al. 1998, PASA, 15, 56

\bibitem[Getman et al.(2005)]{get05} Getman, K.~V., et al.\
2005, \apjs, 160, 319


\bibitem[Gotthelf (2003)]{got03}Gotthelf, E.V. 2003, ApJ, 591, 361


\bibitem[Hessels et al.(2004)]{het04}Hessels, J.W.T., {\it et al.} 2004 ApJ, 612, 389

\bibitem[Kalberla et al.(2005)]{ket05}Kalberla, P.M.W. et al. 2005, A\&A, 440, 775

\bibitem[Kargaltsev et al.(2007)]{kpg07} Kargaltsev, O.,
Pavlov, G.~G., \& Garmire, G.~P.\ 2007, \apj, 660, 1413

\bibitem[Kargaltsev \& Pavlov(2008)]{kp08} Kargaltsev, O.,
\& Pavlov, G.~G.\ 2008, ArXiv e-prints, 801, arXiv:0801.2602

\bibitem[Kennel \& Coroniti(1984)]{kc84} Kennel, C.~F., \&
Coroniti, F.~V.\ 1984, \apj, 283, 694

\bibitem[Kenyon \& Hartmann(1995)]{kah95} Kenyon, S.~J., \& Hartmann, L.\ 1995, \apjs, 101, 117

\bibitem[Komissarov \& Lyubarsky(2004)]{kl04} Komissarov, S. \& Lyubarsky, Y.\ 2004, Ap\&SS, 293, 107



%
\bibitem[Mori et al.(2001)]{met01}Mori, K. et al 2001, ASPC, 251, 576

\bibitem[Ng \& Romani(2004)]{nr04} Ng, C.-Y., \& Romani,
R.~W.\ 2004, \apj, 601, 479


\bibitem[Ng \& Romani(2008)]{nr08}Ng, C.-Y., \& Romani, R.W.\ 2008, \apj, 673, 411


\bibitem[Pavlov et al.(2001)]{pet01}Pavlov, G. G., Zavlin, V.E., Sanwal, D., Burwitz, V., \& Garmire, G.P. 2001, ApJ, 552, L129

\bibitem[Pavlov et al.(2003)]{pet03}Pavlov, G.G., Teter, M.A., Kargaltsev, O.
\& Sanwal, D., 2003, ApJ, 591, 1157

\bibitem[Pavlov et al.(2007)]{pet07} Pavlov, G.~G.,
Kargaltsev, O., \& Brisken, W.~F.\ 2007, ArXiv e-prints, 707,
arXiv:0707.3529

\bibitem[Roberts, Romani, \& Kawai (2001)]{rrk01} Roberts, M.~S.~E.,
Romani, R.~W., \& Kawai, N.\ 2001, \apjs, 133, 451

\bibitem[Roberts et al.(2002)]{ret02} Roberts, M.~S.~E.,
Hessels, J.~W.~T., Ransom, S.~M., Kaspi, V.~M., Freire, P.~C.~C., Crawford,
F., \& Lorimer, D.~R.\ 2002, \apjl, 577, L19

\bibitem[Rees \& Gunn(1974)]{rg74} Rees, M.~J., \& Gunn,
J.~E.\ 1974, \mnras, 167, 1


\bibitem[Reynolds \& Chevalier(1984)]{rc84} Reynolds, S.P.,
\& Chevalier, R.~A.\ 1984, \apj, 278, 630



\bibitem[Romani \& Yadigaroglu(1995)]{ry95} Romani, R.~W.,
\& Yadigaroglu, I.-A.\ 1995, \apj, 438, 314

\bibitem[Romani et al.(2005)]{ret05}Romani, R. W., Ng, C.-Y., Dodson, R., \& Brisken, W. 2005, ApJ, 631, 480

\bibitem[Rosse(1844)]{ros44}Rosse 1844, Phil. Trans. R.S., 321.

\bibitem[Schlegel et al.(1998)]{sch98} Schlegel, D.~J., Finkbeiner, D.~P., \& Davis, M.\ 1998, \apj, 500, 525


\bibitem[van der Swaluw et al.(2000)]{vds00} van der Swaluw,
E., Achterberg, A., Gallant, Y.~A., \& T{\'o}th, G.\ 2000, ArXiv
Astrophysics e-prints, arXiv:astro-ph/0012440

\bibitem[van der Swaluw, Downes \& Keegan(2004)]{vdsdk04}van der Swaluw, E., Downes, T.P \& Keegan, R. 2004, A\&A, 420 937

\bibitem[Zavlin et al.(1996)]{zet96} Zavlin V.E. et al., 1996, A\&A 315, 141

\end{thebibliography}
\end{document}